\documentclass[conference,twoside]{IEEEtran}
\usepackage[T1]{fontenc}
\usepackage{booktabs}
\usepackage{graphicx}
\usepackage[hidelinks]{hyperref}
\usepackage{float}
\usepackage{microtype}
\title{An Empirical Evaluation of Locally Deployed LLMs for Bug Detection in Python Code}

\author{\IEEEauthorblockN{Jelena Ili\'{c} Vuli\'{c}evi\'{c}}
\IEEEauthorblockA{Independent Researcher \\
Master of Electrical and Computer Engineering \\
Email: jelena.ilic094@gmail.com}}

\begin{document}
\setcounter{page}{1}

\maketitle
\thispagestyle{plain}

\begin{abstract}
Large language models (LLMs) have demonstrated strong performance on a wide range of software engineering tasks, including code generation and analysis. However, most prior work relies on cloud-based models or specialized hardware, limiting practical applicability in privacy-sensitive or resource-constrained environments.

In this paper, we present a systematic empirical evaluation of two locally deployed LLMs, LLaMA 3.2 and Mistral, for real-world Python bug detection using the BugsInPy benchmark. We evaluate 349 bugs across 17 projects using a zero-shot prompting approach at the function level and an automated keyword-based evaluation framework.

Our results show that locally executed models achieve accuracy between 43\% and 45\%, while producing a large proportion of partially correct responses that identify problematic code regions without pinpointing the exact fix. Performance varies significantly across projects, highlighting the importance of codebase characteristics.

The results demonstrate that local models can identify a meaningful share of bugs, though precise localization remains difficult for locally executed LLMs, particularly when handling complex and context dependent bugs in realistic development scenarios.
\end{abstract}

\section{Introduction}

Software bugs remain a major challenge in modern software development. Identifying and fixing bugs consumes a substantial portion of development time, while undetected errors can lead to serious failures in production systems. A software bug can be defined as an unintended error in source code that results in incorrect or unexpected behavior, ranging from minor issues to critical system failures. Consequently, there is a strong need for automated techniques that can assist developers in detecting and resolving such problems.

Recent advances in large language models (LLMs), which are neural networks trained on large-scale text and code corpora, have demonstrated promising capabilities in code understanding and generation. Models such as GPT-4~\cite{achiam2023} and Claude~\cite{anthropic2024} have shown strong performance on tasks related to bug detection. However, these models are typically accessed through cloud based services, requiring internet connectivity and incurring usage costs. In addition, sending proprietary code to external servers raises important concerns related to privacy and data security.

Locally executed models offer an alternative approach. Unlike cloud based solutions, these models run entirely on local hardware, ensuring data privacy and eliminating usage costs. Recent progress in model efficiency has made it feasible to run capable LLMs on consumer grade hardware. Models such as LLaMA 3.2 (8B)~\cite{llama32} and Mistral 7B~\cite{jiang2023} demonstrate competitive performance despite their relatively smaller size, making them suitable candidates for local deployment in practical settings.

Despite these advances, there is limited work that systematically evaluates the effectiveness of locally deployed LLMs for bug detection in real world software projects. Most existing studies focus on large cloud based models, rely on simplified benchmarks, or target specific categories such as security vulnerabilities. As a result, the ability of smaller locally executable models to handle realistic debugging scenarios remains insufficiently understood.

The Python programming language provides a suitable context for this evaluation. Python is widely used across domains such as web development, data science, and machine learning~\cite{tiobe2024}. At the same time, its dynamic typing and flexible syntax can make certain types of bugs difficult to detect using traditional static analysis techniques. In contrast, LLMs can reason about program behavior based on semantic patterns without relying on explicit type information. The availability of the BugsInPy dataset~\cite{widyasari2020}, which contains real bugs collected from well known Python projects, further enables a rigorous empirical evaluation.

To fill this gap, we evaluate two locally executed LLMs, LLaMA 3.2 and Mistral, on the BugsInPy dataset. The dataset includes real bugs from 17 Python projects such as pandas, keras, matplotlib, and scrapy. We adopt a zero-shot prompting approach, where models are queried without task specific examples, and evaluate their responses at the function level by analyzing individual functions instead of complete files.

The main contributions of this paper are as follows:

\begin{itemize}
\item We present a systematic empirical evaluation of two locally executed LLMs on a dataset of real world Python bugs.
\item We design and implement an automated keyword based evaluation method for assessing free text bug detection responses.
\item We analyze performance variation across multiple Python projects and identify factors that influence detection accuracy.
\item We provide all code and data required to reproduce our experiments.
\end{itemize}

\section{Related Work}

Large Language Models (LLMs) have recently emerged as a prominent research direction in software engineering, with applications in code generation, bug detection, and automated debugging.

Chen et al.~\cite{chen2021codex} introduced Codex, a GPT-based model fine-tuned on publicly available code, evaluating its ability to generate functional Python programs from natural language descriptions using the HumanEval benchmark. Feng et al.~\cite{codebert} proposed CodeBERT, trained jointly on natural language and programming language data, enabling tasks such as code search and documentation generation. Both approaches demonstrate strong performance on code understanding but focus on generation rather than detection of bugs in realistic settings.

Bug detection and fault localization have received growing attention in subsequent work. Kang et al.~\cite{kang2023fault} proposed AutoFL, which uses LLMs for fault localization while generating natural language explanations alongside predictions, improving interpretability in large codebases. A related effort by Mhatre et al.~\cite{llm_guard_2025} evaluated cloud-based models on a benchmark spanning both Python and C++, finding that defect complexity is the primary factor governing detection accuracy. These studies converge on the observation that performance degrades when bugs require cross-function reasoning.

Test quality and structured bug reporting represent a different dimension of the same problem. Santana et al.~\cite{santana2025test} measured how well LLMs detect and refactor test smells, reporting variation across model families and task complexity. Acharya and Ginde~\cite{acharya2025bugreport} applied instruction-tuned models to convert unstructured bug reports into structured form, finding that open-weight models can approach proprietary system performance on this task, though results vary across report types.

Limitations in handling long contexts and complex program logic have been studied directly. Lee et al.~\cite{bug_stack_2024} examined large Python codebases and documented consistent performance degradation as code complexity grows. Tambon et al.~\cite{tambon2024bugs} analyzed bugs produced by LLMs themselves, proposing a taxonomy of recurring failure patterns in generated code. Both studies highlight that current models remain sensitive to context length and structural complexity.

Benchmark construction has also been a productive research direction. Widyasari et al.~\cite{widyasari2020} released BugsInPy, a curated set of real Python bugs with reproducible test cases drawn from well-known open-source projects. Aguilar et al.~\cite{aguilar2023} later identified reproducibility issues in this dataset and proposed revisions to support more reliable evaluation. Pushkar et al.~\cite{pushkar2025} extended evaluation to multi-vulnerability settings, showing that model performance drops systematically as the number of co-occurring defects increases. Zhang et al.~\cite{zhang2025apr} contributed a systematic review of LLM-based automated program repair, categorizing methods and identifying open challenges.

The availability of open-weight models such as LLaMA~\cite{llama} and Mistral~\cite{jiang2023} has made local deployment a practical option, eliminating cloud dependencies and usage costs. Despite this, prior work has largely evaluated large cloud-hosted models, leaving the capabilities of consumer-grade offline deployments inadequately characterized. This study evaluates LLaMA 3.2 and Mistral in a fully offline setting on the BugsInPy benchmark, providing an empirical basis for assessing their utility in realistic, resource-constrained debugging scenarios.

\section{Methodology}
\label{sec:methodology}

\subsection{Dataset}
We use the BugsInPy dataset~\cite{widyasari2020}, which contains 501 real bugs collected from 17 well known open source Python projects, including pandas, keras, matplotlib, scrapy, ansible, and others. These projects cover a diverse set of domains such as data science, web development, machine learning, and developer tools, providing a representative sample of real world Python code. Each bug instance includes a patch file in unified diff format indicating the modified lines, metadata with buggy and fixed commit identifiers, a failing test case that exposes the bug, and scripts for reproducing the execution environment.

We attempt to process all 501 bugs in the dataset. A bug is excluded if one of the following conditions is met: (1) the original source file cannot be retrieved due to repository changes or unavailable commits, (2) the buggy line cannot be located in the retrieved file, or (3) no enclosing \texttt{def} or \texttt{class} statement can be identified to extract a complete function. After applying these criteria, 349 bugs are successfully processed and used in the evaluation, corresponding to 69.7\% of the dataset. The distribution of processed bugs across projects is shown in Table~\ref{tab:bugs_per_project}.

\begin{table}
\centering
\caption{Number of processed bugs per project from the BugsInPy dataset}
\label{tab:bugs_per_project}
\begin{tabular}{lc}
\toprule
Project & Bugs Used \\
\midrule
pandas       & 96 \\
scrapy       & 32 \\
thefuck      & 29 \\
luigi        & 28 \\
youtube-dl   & 34 \\
keras        & 31 \\
black        & 19 \\
matplotlib   & 19 \\
ansible      & 13 \\
fastapi      & 13 \\
tornado      & 10 \\
tqdm         & 7  \\
spacy        & 5  \\
httpie       & 4  \\
sanic        & 4  \\
cookiecutter & 3  \\
PySnooper    & 2  \\
\midrule
Total        & 349 \\
\bottomrule
\end{tabular}
\end{table}

\subsection{Code Extraction}
For each bug, we extract three elements from the patch file: the file path, the buggy lines, and the corresponding fixed lines. The original source file is then retrieved using the buggy commit identifier through the GitHub raw content API, ensuring that the exact version of the code is obtained.

From the retrieved file, we identify and extract the function that contains the buggy line. This is done by scanning upward to locate the nearest \texttt{def} or \texttt{class} statement that marks the beginning of the function, and scanning downward to identify the next such statement, which marks its end. This procedure captures the full function, including its signature, documentation, and implementation.

Function level extraction is performed based on preliminary experiments with alternative approaches. Providing the entire file often leads models to produce general summaries instead of focusing on the bug, as the relevant lines are diluted within a larger context. In contrast, providing only the buggy line with limited surrounding context does not offer sufficient information for understanding the intended behavior. Extracting the full function provides a balance between context and focus. In cases where multiple buggy lines are present, the first occurrence is used as an anchor for function extraction.

\subsection{Prompt Design}
Each model is queried using the same zero-shot prompt format:

\begin{verbatim}
Here is a Python function.
It contains a bug.
Find the bug and explain how to fix it.
\end{verbatim}

A zero-shot setting is used for two reasons. First, it reflects a realistic usage scenario in which developers interact with models without task specific configuration. Second, it allows evaluation of the model’s inherent capability for bug detection without influence from example based prompting.

The prompt explicitly states that the function contains a bug. This design focuses the evaluation on the model’s ability to identify and explain the bug, rather than on binary classification of correctness, and avoids trivial responses that always predict the absence of errors. The function code is appended directly after the prompt.

\subsection{Response Evaluation}
Model outputs are evaluated automatically using a keyword based approach. The evaluation consists of three steps.

First, for each bug, we extract keywords from the fixed lines that do not appear in the buggy version. These keywords typically correspond to new identifiers, function calls, or concepts introduced by the fix.

Second, we remove common English words and tokens shorter than four characters to reduce noise and avoid accidental matches. We note that the evaluation approach is heuristic and may not fully capture semantic correctness. In particular, keyword matching may produce false positives when keywords appear as substrings, and false negatives when correct answers use alternative terminology. Therefore, the reported results should be interpreted as an approximation of model performance rather than an exact measure.

Third, each response is assigned one of three labels. A response is considered \textit{correct} if it includes at least one keyword associated with the fix. A response is labeled \textit{partial} if it does not include keywords but indicates the presence of a problem using terms such as bug, issue, or error. Otherwise, the response is labeled \textit{wrong}.

This evaluation approach is conservative. Correct responses that use different terminology may not match the extracted keywords and can therefore be assigned a lower score. As a result, the reported accuracy should be interpreted as a lower bound of the true performance. 

In addition to fully correct predictions, some responses are labeled as "partial" when they identify the general location or nature of the bug but do not provide a precise or complete explanation.

To assess the reliability of the automated evaluation, we manually inspected a random sample of 50 model responses. The manual evaluation showed a high level of agreement with the keyword-based labels, indicating that the automated method provides a reasonable approximation of model performance despite its limitations.

For the purpose of statistical evaluation, we map model outputs to binary outcomes. Responses labeled as "correct" are treated as correct predictions, while "partial" and other responses are treated as incorrect.

\subsection{Models and Hardware}
We evaluate two locally executed LLMs using the Ollama framework~\cite{ollama2023}, which enables efficient model execution on consumer hardware. The first model is LLaMA 3.2 (8B parameters)~\cite{llama32}, and the second is Mistral (7B parameters)~\cite{jiang2023}. Both are referred to by their short names throughout the remainder of this paper.

All experiments are conducted on a MacBook Pro 14 inch (2021) equipped with an Apple M1 Pro processor and 16GB of unified memory, running macOS Sequoia 15.7.3. The software environment includes Python 3.13 and Ollama version 0.18.0. All experiments are executed fully offline, with no communication with external services.

The average response time per query was approximately 7 seconds for LLaMA 3.2 and 13 seconds for Mistral, resulting in a total experiment runtime of approximately 40 minutes for LLaMA 3.2 and 75 minutes for Mistral. All models were executed using 4-bit quantization (Q4\_K\_M) through the Ollama framework to ensure efficient inference on consumer-grade hardware.

\subsection{Reproducibility}
To ensure reproducibility, all code and experimental results are publicly available.\footnote{\url{https://github.com/insajder/llm-bug-detection}} This includes scripts for dataset processing, source code retrieval, model interaction, and response evaluation. All outputs are stored in JSON format and can be regenerated using a single script. The BugsInPy dataset is also publicly available.\footnote{\url{https://github.com/soarsmu/BugsInPy}}

\subsection{Bug Type Classification}
To enable fine-grained analysis, each of the 349 bugs was manually assigned to one of nine categories based on the nature of the change between the buggy and fixed versions: \textit{Null/None Check}, \textit{Return Value}, \textit{Conditional Logic}, \textit{Indexing}, \textit{Error Handling}, \textit{Loop Logic}, \textit{Type Conversion}, \textit{Comparison Operator}, and \textit{Other/Complex}. Classification was performed by inspecting the modified tokens in each patch without reference to model outputs.

\section{Results}

This section presents the empirical evaluation of LLaMA 3.2 and Mistral on the BugsInPy benchmark. We first report overall performance, followed by a detailed analysis across individual projects and score distributions.

\subsection{Overall Results}
Table~\ref{tab:overall} presents the overall results.

\begin{table}
\caption{Overall Results by Model}
\label{tab:overall}
\centering
\begin{tabular}{lcccc}
\toprule
Model & Correct & Partial & Wrong & Accuracy \\
\midrule
LLaMA 3.2 & 151 & 171 & 27 & 43.3\% \\
Mistral   & 155 & 161 & 33 & 44.4\% \\
\bottomrule
\end{tabular}
\end{table}

\begin{figure}
\centering
\includegraphics[width=\columnwidth]{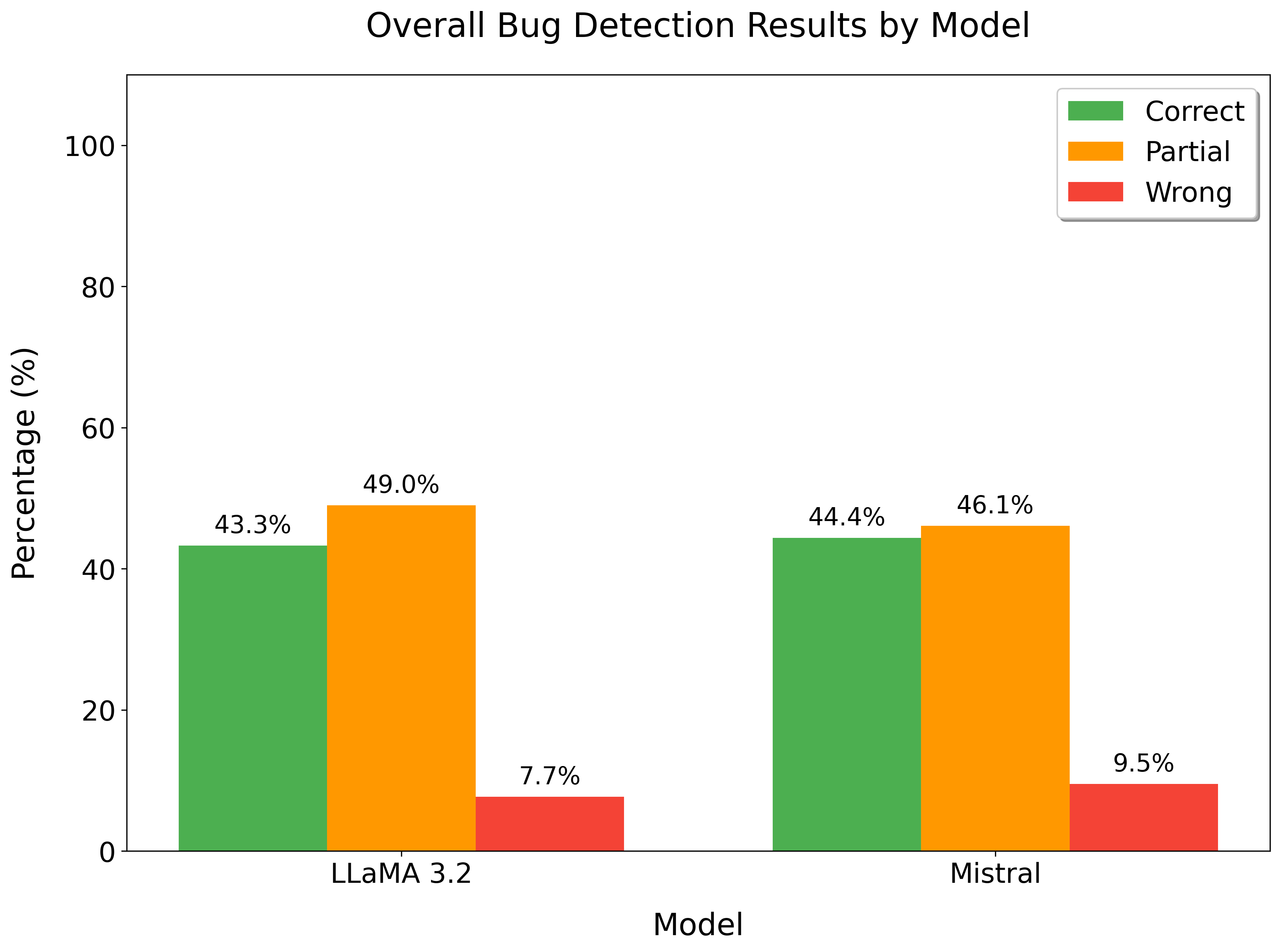}
\caption{Overall bug detection results by model.}
\label{fig:overall}
\end{figure}

As shown in Table~\ref{tab:overall}, Mistral achieves 44.4\% accuracy, slightly outperforming LLaMA 3.2 by 1.1 percentage points (43.3\%). Both models produce a substantial proportion of partial responses, indicating that they are often able to recognize that a problem exists in the code, but struggle to precisely identify the underlying issue and its fix. These results indicate that, in practice, locally deployed LLMs tend to be more reliable at detecting the presence of bugs than at accurately localizing and explaining them. Figure~\ref{fig:overall} illustrates the distribution of correct, partial, and wrong responses for both models.

\subsection{Results by Project}

Table~\ref{tab:byproject} presents the accuracy achieved by each model across individual projects. The results show substantial variation, ranging from 0\% on tqdm to 100\% on PySnooper.

LLaMA 3.2 consistently outperforms Mistral on several projects, including fastapi (by 7.7 percentage points) and spacy (by 20 percentage points). In contrast, Mistral achieves better performance on projects such as ansible (by 7.7 percentage points), pandas (by 6.3 percentage points), and tornado (by 10 percentage points).

\begin{table}
\caption{Accuracy by Project (\%)}
\label{tab:byproject}
\centering
\begin{tabular}{lcc}
\toprule
Project & LLaMA 3.2 & Mistral \\
\midrule
PySnooper    & 100.0 & 100.0 \\
black        & 73.7  & 68.4  \\
fastapi      & 69.2  & 61.5  \\
ansible      & 61.5  & 69.2  \\
keras        & 54.8  & 48.4  \\
httpie       & 50.0  & 50.0  \\
tornado      & 50.0  & 60.0  \\
matplotlib   & 47.4  & 47.4  \\
scrapy       & 40.6  & 43.8  \\
spacy        & 40.0  & 20.0  \\
pandas       & 38.5  & 44.8  \\
youtube-dl   & 38.2  & 35.3  \\
thefuck      & 34.5  & 37.9  \\
cookiecutter & 33.3  & 33.3  \\
luigi        & 28.6  & 28.6  \\
sanic        & 25.0  & 0.0   \\
tqdm         & 0.0   & 14.3  \\
\bottomrule
\end{tabular}
\end{table}

As a simple baseline, random keyword matching would yield negligible accuracy (below 5\%), indicating that both models perform substantially above chance.

To evaluate whether the observed difference between the two models is statistically significant, we applied McNemar's test~\cite{mcnemar1947} on paired predictions. Model outputs were aligned by bug identifier and mapped to binary outcomes (correct vs.\ incorrect). Of 349 paired predictions, both models agreed on 126 correct and 169 incorrect cases. LLaMA 3.2 was correct where Mistral was not in 25 cases, while Mistral was correct where LLaMA 3.2 was not in 29 cases. The exact binomial McNemar's test yields $p = 0.68$, indicating no statistically significant difference between the two models.

\begin{figure}
\centering
\includegraphics[width=\columnwidth]{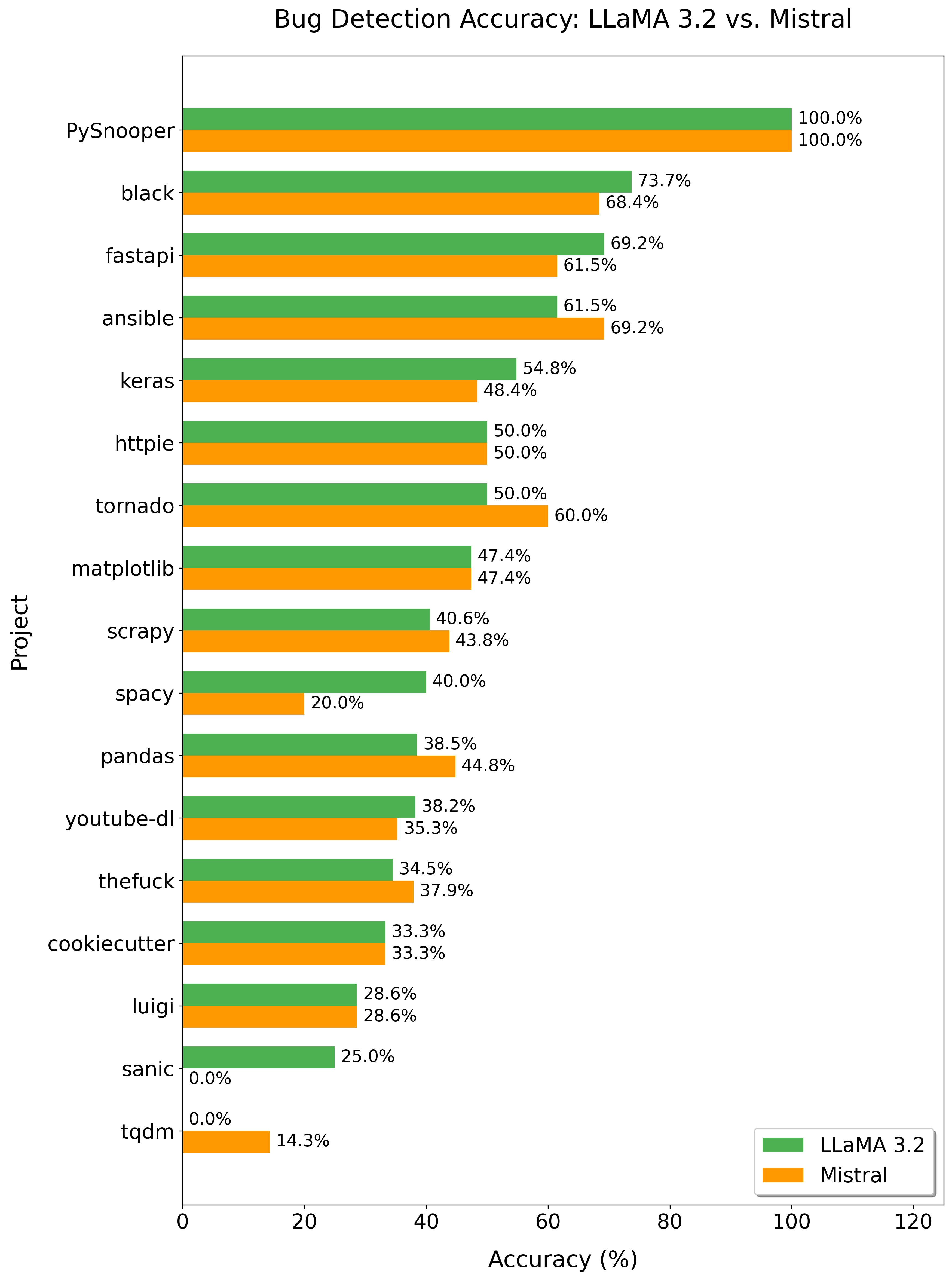}
\caption{Bug detection accuracy by project.}
\label{fig:byproject}
\end{figure}

Figure~\ref{fig:byproject} shows the accuracy of both models across all projects, highlighting substantial variation in performance.

All reported metrics are computed directly from the raw experimental outputs to ensure consistency and reproducibility.

\subsection{Score Distribution by Project}

Table~\ref{tab:distribution} presents the distribution of correct, partial, and wrong responses for both models across all projects.

\begin{table}
\caption{Score Distribution by Project (\%). Rows may not sum to exactly 100 due to rounding.}
\label{tab:distribution}
\centering
\begin{tabular}{lcccccc}
\toprule
& \multicolumn{3}{c}{LLaMA 3.2} & \multicolumn{3}{c}{Mistral} \\
\cmidrule(lr){2-4} \cmidrule(lr){5-7}
Project & Correct & Partial & Wrong & Correct & Partial & Wrong \\
\midrule
PySnooper & 100.0 & 0.0 & 0.0 & 100.0 & 0.0 & 0.0 \\
black & 73.7 & 26.3 & 0.0 & 68.4 & 31.6 & 0.0 \\
fastapi & 69.2 & 30.8 & 0.0 & 61.5 & 38.5 & 0.0 \\
ansible & 61.5 & 38.5 & 0.0 & 69.2 & 30.8 & 0.0 \\
keras & 54.8 & 38.7 & 6.5 & 48.4 & 48.4 & 3.2 \\
tornado & 50.0 & 50.0 & 0.0 & 60.0 & 20.0 & 20.0 \\
httpie & 50.0 & 50.0 & 0.0 & 50.0 & 50.0 & 0.0 \\
matplotlib & 47.4 & 15.8 & 36.8 & 47.4 & 15.8 & 36.8 \\
scrapy & 40.6 & 59.4 & 0.0 & 43.8 & 56.2 & 0.0 \\
spacy & 40.0 & 60.0 & 0.0 & 20.0 & 80.0 & 0.0 \\
pandas & 38.5 & 45.8 & 15.6 & 44.8 & 35.4 & 19.8 \\
youtube-dl & 38.2 & 58.8 & 2.9 & 35.3 & 61.8 & 2.9 \\
thefuck & 34.5 & 65.5 & 0.0 & 37.9 & 62.1 & 0.0 \\
cookiecutter & 33.3 & 66.7 & 0.0 & 33.3 & 66.7 & 0.0 \\
luigi & 28.6 & 71.4 & 0.0 & 28.6 & 67.9 & 3.6 \\
sanic & 25.0 & 50.0 & 25.0 & 0.0 & 75.0 & 25.0 \\
tqdm & 0.0 & 85.7 & 14.3 & 14.3 & 71.4 & 14.3 \\
\bottomrule
\end{tabular}
\end{table}

The results reveal several consistent patterns across projects. Both models achieve perfect performance on PySnooper, with 100\% correct responses and no errors. In contrast, projects such as tqdm show very low correctness, with LLaMA 3.2 achieving 0\% and Mistral 14.3\%, while the majority of responses are classified as partial, indicating that models often recognize the presence of an issue but fail to identify the exact fix.

Projects such as matplotlib exhibit a relatively high proportion of wrong responses for both models (36.8\%), suggesting that bugs in these codebases are more difficult to diagnose accurately. Similarly, pandas and sanic show a non-negligible share of wrong predictions, particularly for Mistral.

On the other hand, projects such as ansible, black, and fastapi are dominated by correct and partial responses, with very few or no wrong predictions, indicating that bugs in these projects are easier to localize and interpret.

\begin{figure}
\centering
\includegraphics[width=\columnwidth]{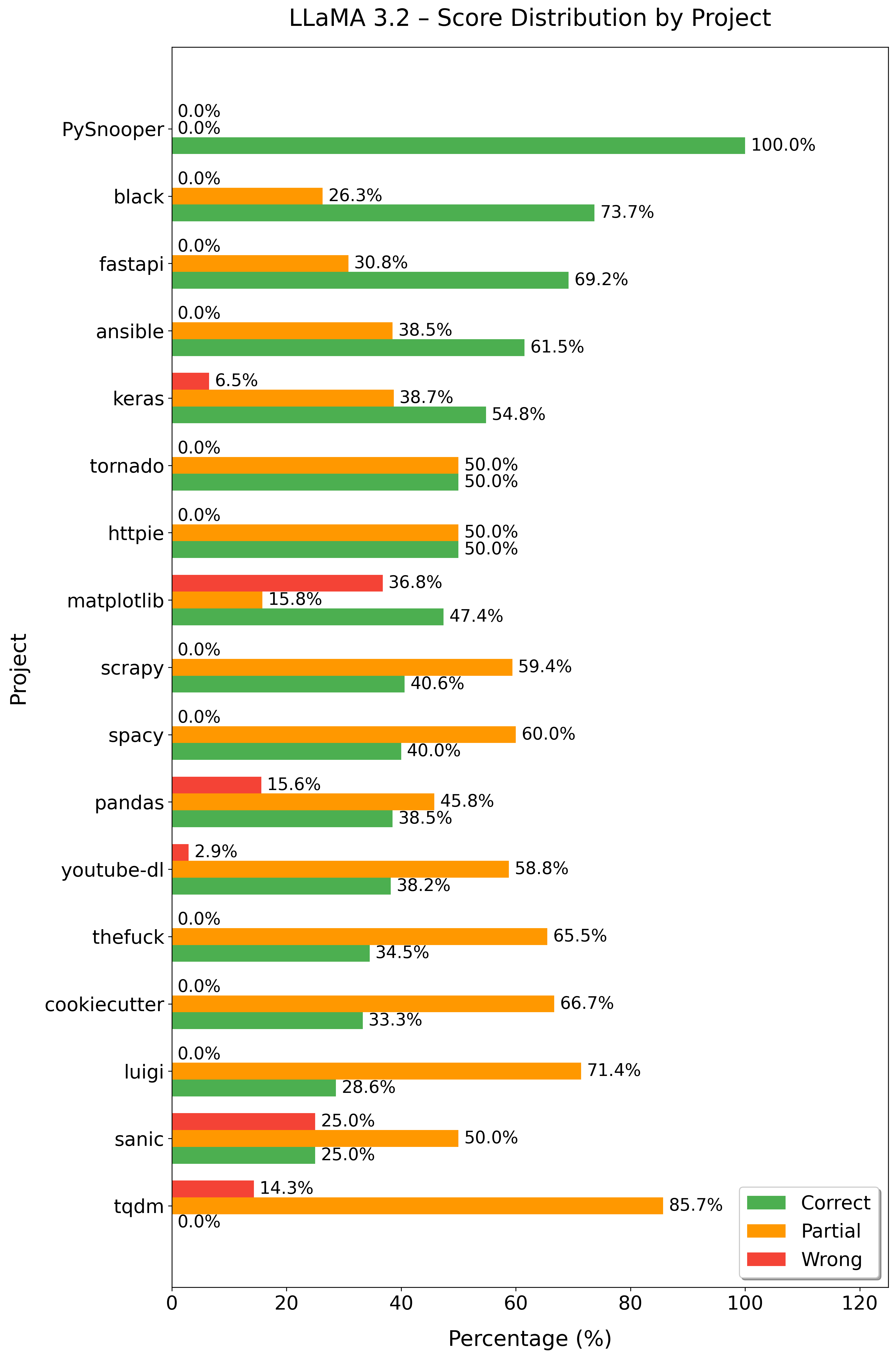}
\caption{Score distribution by project for LLaMA 3.2.}
\label{fig:distribution_llama}
\end{figure}

\begin{figure}
\centering
\includegraphics[width=\columnwidth]{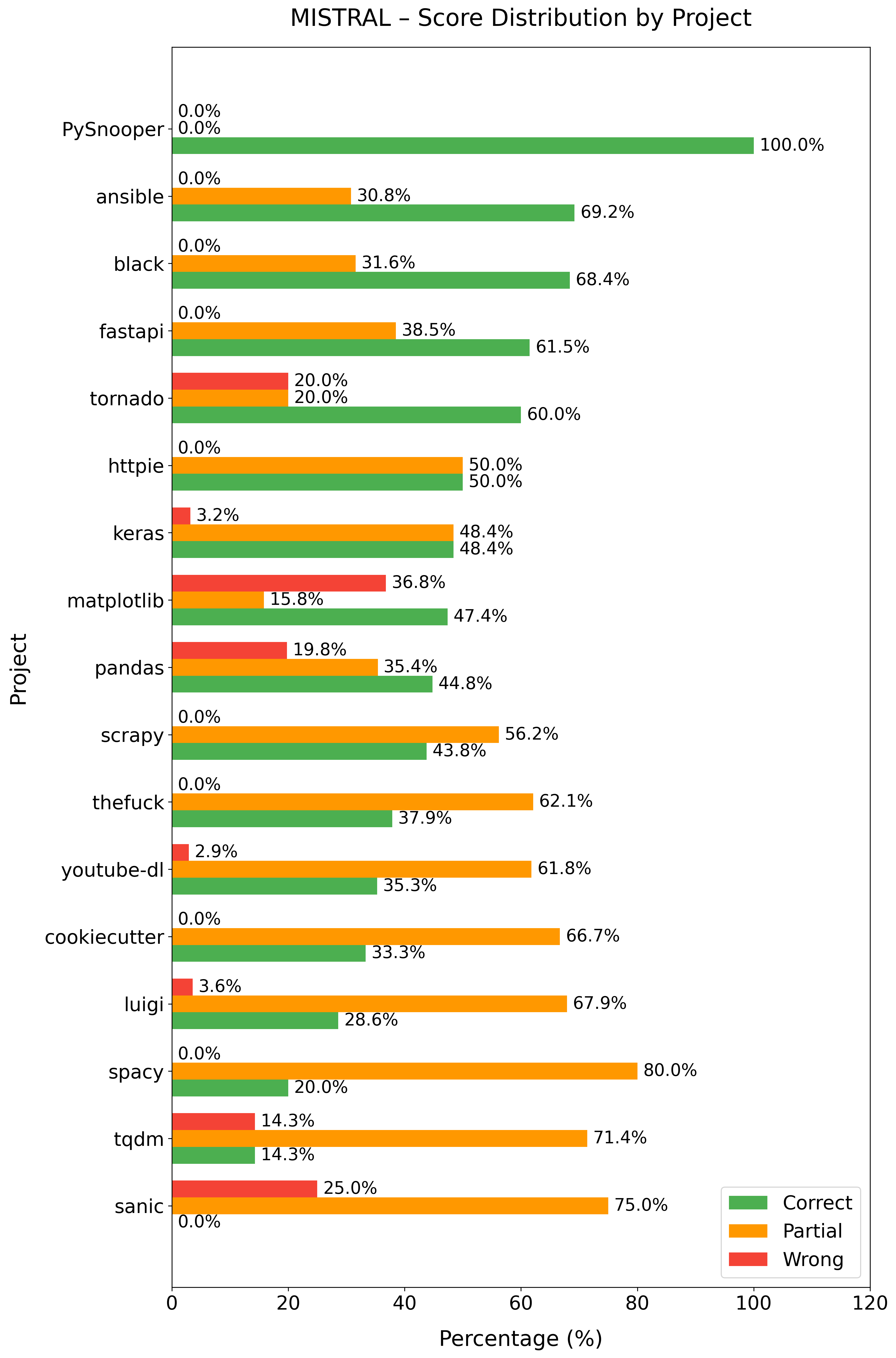}
\caption{Score distribution by project for Mistral.}
\label{fig:distribution_mistral}
\end{figure}

Figures~\ref{fig:distribution_llama} and \ref{fig:distribution_mistral} provide a visual comparison of score distributions across all projects for each model. LLaMA 3.2 achieves strong performance on projects such as black (73.7\%) and fastapi (69.2\%), while also exhibiting a high proportion of partial responses on projects like luigi (71.4\%) and tqdm (85.7\%). Mistral shows comparable performance on several projects, outperforming LLaMA 3.2 on ansible (69.2\% vs 61.5\%) and tornado (60.0\% vs 50.0\%), but also produces more partial and wrong responses on projects such as spacy (80.0\% partial) and sanic (25.0\% wrong).

All reported metrics are computed directly from the raw experimental outputs to ensure consistency and reproducibility.

\subsection{Results by Bug Type}

To better understand model behavior, we categorize the 349 evaluated bugs into nine types based on the nature of the change between the buggy and fixed versions. Categories are derived by inspecting the modified tokens in each patch: \textit{Null/None Check}, \textit{Return Value}, \textit{Conditional Logic}, \textit{Indexing}, \textit{Error Handling}, \textit{Loop Logic}, \textit{Type Conversion}, \textit{Comparison Operator}, and \textit{Other/Complex} for cases involving multiple interacting components.

Table~\ref{tab:bugtype} presents the accuracy achieved by each model across bug categories, ordered by frequency.

\begin{table}
\caption{Accuracy by Bug Type (\%)}
\label{tab:bugtype}
\centering
\begin{tabular}{lrcc}
\toprule
Bug Type & Count & LLaMA 3.2 & Mistral \\
\midrule
Null/None Check      & 79 & 59.5 & 60.8 \\
Return Value         & 78 & 51.3 & 51.3 \\
Conditional Logic    & 55 & 40.0 & 36.4 \\
Indexing             & 44 & 36.4 & 47.7 \\
Error Handling       & 16 & 50.0 & 62.5 \\
Loop Logic           & 11 & 36.4 & 45.5 \\
Comparison Operator  &  2 & 50.0 & 50.0 \\
Type Conversion      &  4 &  0.0 &  0.0 \\
Other/Complex        & 60 & 21.7 & 16.7 \\
\bottomrule
\end{tabular}
\end{table}

Across bug categories, consistent differences emerge. Both models perform best on \textit{Null/None Check} bugs, achieving around 60\% accuracy, which reflects the syntactic regularity of such errors, missing guard clauses or incorrect \texttt{None} comparisons tend to follow recognizable patterns that LLMs can match against training data. \textit{Return Value} bugs also yield competitive accuracy near 51\% for both models, as incorrect return expressions are often localized within a single line and do not require cross-function reasoning.

Performance drops considerably for \textit{Conditional Logic} and \textit{Indexing} bugs, where correct detection requires understanding of program state and data shape. \textit{Type Conversion} bugs prove the most difficult, with both models scoring 0\%, likely because such errors manifest only at runtime and depend on implicit type contracts that are not apparent from the function body alone.

The \textit{Other/Complex} category, which captures bugs involving interactions between multiple components, yields the lowest accuracy after type conversion (21.7\% for LLaMA~3.2, 16.7\% for Mistral). This aligns with the function-level analysis limitation noted in Section~\ref{sec:methodology}: when a bug's root cause spans multiple functions or relies on shared state, a single-function prompt does not provide sufficient context for reliable detection.

Notably, Mistral outperforms LLaMA~3.2 on \textit{Indexing} (+11.3 pp) and \textit{Error Handling} (+12.5 pp) bugs, while LLaMA~3.2 shows a slight advantage on \textit{Conditional Logic} (+3.6 pp). These differences suggest complementary strengths, and motivate future work on ensemble or model selection strategies.

\begin{figure}
\centering
\includegraphics[width=\columnwidth]{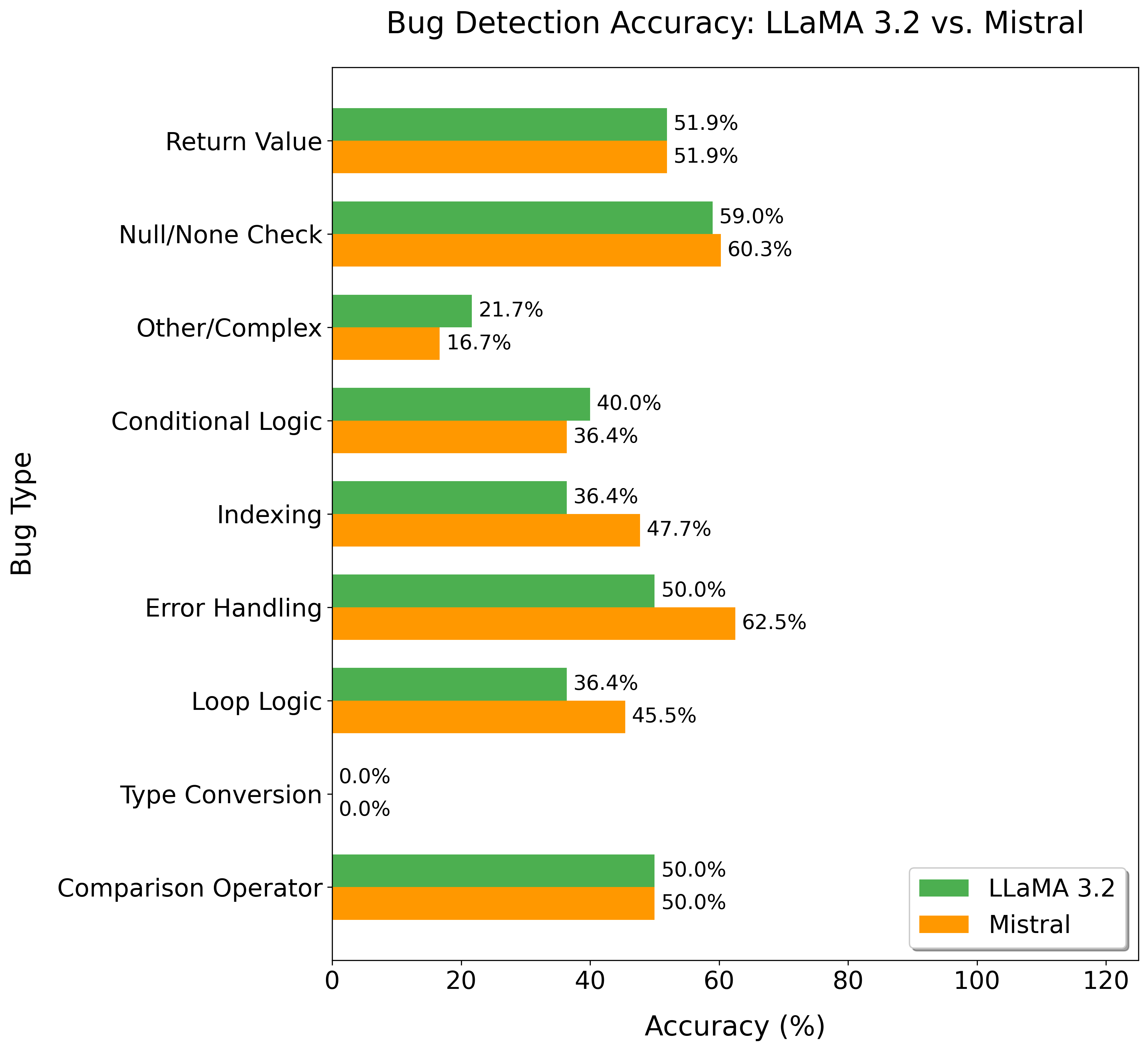}
\caption{Bug detection accuracy by bug type for both models. Numbers in parentheses indicate the count of bugs per category.}
\label{fig:bugtype}
\end{figure}

Figure~\ref{fig:bugtype} summarizes these results visually.

All reported metrics are computed directly from the raw experimental outputs to ensure consistency and reproducibility.

\section{Discussion}

\subsection{Interpretation of Results}

The results indicate that locally deployed LLMs are capable of identifying a substantial portion of real-world Python bugs, achieving accuracy between 43\% and 45\%. While this level of performance is not sufficient for fully automated debugging, it is significant given that both models operate entirely on consumer-grade hardware without GPU acceleration, external APIs, or internet connectivity. Removing both the cost and connectivity requirements makes such systems accessible to individual developers and resource-constrained teams.

Importantly, the BugsInPy dataset consists of real defects extracted from production-level open-source projects, which are considerably more challenging than synthetic benchmarks. The fact that local models correctly identify over 40\% of such bugs suggests that they can already provide meaningful assistance in practical debugging workflows.

A key observation is the high proportion of partial responses, which account for nearly half of all outputs. In these cases, models successfully identify that a function contains problematic logic but fail to precisely locate the root cause or propose an exact fix. Although not fully correct, such responses remain practically useful, as they narrow the search space and highlight suspicious regions of code.

\subsection{Comparative Performance}

Although the overall accuracy difference between LLaMA 3.2 and Mistral is relatively small (approximately 1 percentage point), the results reveal substantial variation across projects. This indicates that model effectiveness is highly dependent on codebase characteristics rather than global model superiority.

LLaMA 3.2 tends to perform better on projects with more structured and conventional code patterns, while Mistral shows competitive or superior performance on projects involving less predictable logic. These results suggest that the models exhibit complementary strengths, and that combining multiple models may be a promising direction for improving robustness.

\subsection{Failure Cases}

The lowest performance is observed on projects such as \texttt{tqdm}, \texttt{sanic}, \texttt{luigi}, and \texttt{spacy}, where accuracy drops significantly. These projects typically involve complex abstractions, implicit dependencies, or interactions across multiple components.

In such cases, understanding the bug requires broader contextual information beyond a single function, including inter-function dependencies and framework-specific behavior. This highlights a fundamental limitation of function-level analysis and suggests that incorporating additional context may be necessary for more accurate bug detection.

\subsection{Threats to Validity}

The primary threat to validity lies in the keyword-based evaluation approach. While efficient and fully automated, this method provides only an approximate measure of correctness and may both underestimate and overestimate true model performance. In particular, semantically correct responses that use different terminology may be misclassified as partial or incorrect, while coincidental keyword matches may lead to false positives.

As a result, the reported accuracy should be interpreted as a conservative lower bound rather than an exact measure of model capability. Incorporating manual evaluation or semantic similarity metrics would provide a more reliable assessment.

Additional threats include the limited scope of the evaluation. Although BugsInPy is widely used, it does not cover all programming styles or domains, and only two models are evaluated in this study. Expanding the analysis to additional datasets and models would improve the generalizability of the findings. Another limitation is that the prompt explicitly states that the function contains a bug. This may inflate model performance compared to real-world scenarios where the presence of a bug is not known in advance.

\subsection{Practical Implications}

Even so, the results show that locally deployed LLMs can serve as effective first-pass debugging assistants. Rather than replacing developers, they are most useful for identifying suspicious code regions and providing initial diagnostic insights.

The offline nature of local deployment is particularly valuable in privacy-sensitive environments, where source code cannot be shared externally. In addition, the absence of API costs makes such systems accessible to a wider range of users, including students, independent developers, and small teams.

\section{Conclusion}

\subsection{Summary of Findings}

In this paper, we presented a systematic empirical evaluation of locally deployed LLMs for real-world Python bug detection using the BugsInPy dataset. We evaluated two models, LLaMA 3.2 and Mistral, on 349 real bugs across 17 projects using a zero-shot prompting approach and automated evaluation.

The results show that locally executed models achieve accuracy between 43\% and 45\%, while a large proportion of responses remain partially correct, indicating strong detection capability but limited precision in localization and repair.

Performance varies significantly across projects, confirming that codebase characteristics play a critical role in model effectiveness. While both models perform well on structured code, they struggle in scenarios requiring broader context or complex reasoning.

\subsection{Future Work}

Future work should focus on improving both evaluation methodology and model performance. Incorporating manual or semantically aware evaluation would provide more accurate estimates of true capability.

Integrating richer context, such as multiple functions or project-level information, stands as a promising direction from a modeling perspective. Furthermore, the development of retrieval-augmented approaches and hybrid systems could significantly enhance the overall robustness of the bug detection process.

Additional improvements may also be achieved through few-shot prompting or alternative inference strategies. Expanding the evaluation to additional models, datasets, and programming languages would further strengthen the generality of the findings.

Overall, these results demonstrate that meaningful AI-assisted debugging is already feasible on widely available consumer hardware.

\end{document}